\documentclass{article}

\usepackage[preprint]{neurips_2025}

\usepackage[utf8]{inputenc} 
\usepackage[T1]{fontenc}    
\usepackage{hyperref}       
\usepackage{url}            
\usepackage{booktabs}       
\usepackage{amsfonts}       
\usepackage{graphicx}       
\usepackage{nicefrac}       
\usepackage{microtype}      
\usepackage{xcolor}         
\usepackage{amsmath}

\usepackage{arydshln}
\usepackage{booktabs}
\usepackage{multirow}
\usepackage[normalem]{ulem}
\usepackage{graphicx}
\usepackage{pifont}
\usepackage{subcaption} 
\usepackage{dsfont}

\title{LensCraft: Your Professional Virtual Cinematographer}

\author{%
  Zahra Dehghanian \\
  \texttt{zahra.dehghanian97@sharif.edu} \\
  \And
  Morteza Abolghasemi \\
  \texttt{abolghasemi@sharif.edu} \\
  \And
  Hossein Azizinaghsh \\
  \texttt{azizinaghsh@sharif.edu} \\
  \And
 Amir Vahedi\\
  \texttt{amir.vahedi123@sharif.edu} \\
  \And
  Hamid Beigy \\
  \texttt{beigy@sharif.edu} \\
  \And
  Hamid R. Rabiee \\
  \texttt{rabiee@sharif.edu} \\
}

\begin{document}
\maketitle

\begin{abstract}
  Digital creators, from indie filmmakers to animation studios, face a persistent bottleneck: translating their creative vision into precise camera movements. Despite significant progress in computer vision and artificial intelligence, current automated filming systems struggle with a fundamental trade-off between mechanical execution and creative intent. Crucially, almost all previous works simplify the subject to a single point—ignoring its orientation and true volume—severely limiting spatial awareness during filming. LensCraft solves this problem by mimicking the expertise of a professional cinematographer, using a data-driven approach that combines cinematographic principles with the flexibility to adapt to dynamic scenes in real time. Our solution combines a specialized simulation framework for generating high-fidelity training data with an advanced neural model that is faithful to the script while being aware of the volume and dynamic behavior of the subject. Additionally, our approach allows for flexible control via various input modalities, including text prompts, subject trajectory and volume, key points, or a full camera trajectory, offering creators a versatile tool to guide camera movements in line with their vision. Leveraging a lightweight real time architecture , LensCraft achieves markedly lower computational complexity and faster inference while maintaining high output quality. Extensive evaluation across static and dynamic scenarios reveals unprecedented accuracy and coherence, setting a new benchmark for intelligent camera systems compared to state-of-the-art models. 
  Extended results, the complete dataset, simulation environment, trained model weights, and source code are publicly accessible on \href{https://lens-craft.pages.dev//
}{LensCraft Webpage}.

\end{abstract}

\section{Introduction}
The language of cinematography extends far beyond simple camera operations, encompassing a rich representation of visual storytelling elements that can express distinct styles, perceived motions, emotional resonances, and visual perceptions \cite{wang2023jaws}. While this sophisticated language holds the capacity to translate a director's narrative vision into precise camera movements - just as it does in real-world filmmaking - it demands exact alignment and precise execution from automated systems. Each camera movement must not merely follow rules, but rather embody the director's intent with the finesse of an experienced professional cinematographer, whether capturing the subtle tension in a dramatic scene or the dynamic energy of an action sequence \cite{dehghanian2025camera}.

Current automated approaches fall into two limiting categories. Traditional methods that sacrifice creative potential for predictability, reducing the rich language of cinematography to mechanical executions of predefined rules, constraints and objectives \cite{gebhardt2018optimizing,jiang2024cinematic}. On the other hand, more recent learning-based approaches, while capable of generating complex camera sequences, often get lost in this complexity due to imbalanced training data distributions and insufficient richness in their learning examples \cite{courant2025exceptional,bonatti2021batteries}. As a result, these methods struggle to maintain professional quality while staying faithful to user instructions and produce sophisticated sequences that drift from the intended cinematographic direction. This fundamental disconnect becomes particularly apparent when handling diverse filming scenarios, where the balance between professional execution and precise alignment with user intent becomes crucial.

Additionally, one crucial aspect that prior models often overlook is the impact of subject volume. In real-world filmmaking, the scale and physical presence of the subject—whether filming from an ant’s perspective or capturing a building—significantly influence the camera’s behavior and trajectory. However, most of prior works  simplify the subject by representing it as a point \cite{jiang2024cinematic,courant2025exceptional} or by modeling it in a fixed specific form, such as SMPL \cite{wang2024dancecamera3d,chen2024dreamcinema}. This oversimplification, due to limitations in dataset complexity, fails to capture the true dynamics of subject volume, which is essential for generating realistic and context-aware camera movements.

This fundamental disconnect motivated us to develop LensCraft, a novel approach that bridges the gap between professional cinematographic execution and precise user control. In addressing this challenge, we first identified two key obstacles: the absence of a unified, comprehensive standard for a representative cinematographic language, and the lack of a balanced, large-scale, volume-aware dataset that covers the full spectrum of camera movements. To overcome these challenges, we designed a standard cinematography language (SCL) to provide a unified foundation for our approach and developed first fully open-source, carefully designed simulation framework to generate a diverse and well-balanced dataset, serving as the foundational pillars required for our model.

\begin{figure}
  \centering
  \includegraphics[width=0.8\textwidth]{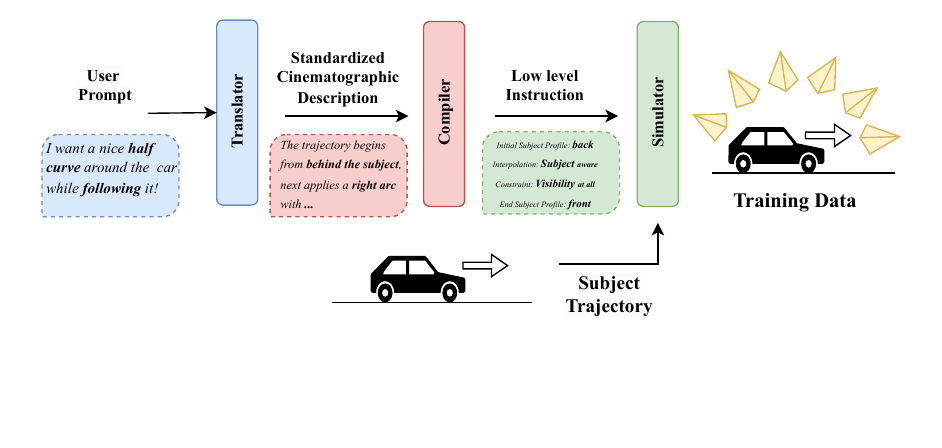}
  \vspace{-45pt}
  \caption{The data generation pipeline illustrating the hierarchical transformation of user intent into camera trajectories in simulation.}
  \label{fig:teaser}
\end{figure}

Leveraging this backbone to combine the solidity of traditional rule-based methods with the creativity of neural models, we designed a model that inherits both features. We developed a lightweight transformer-based architecture that learns and generates complex simulated movements, using only a text prompt or optional key frames, enabling it to produce robust and creatively aligned outputs. Also, to accommodate a wider range of user instructions, we incorporate a Appendix that converts any arbitrary user prompts into our standardized cinematographic description. The data flow and training procedure of our model is shown in Figure \ref{fig:teaser}. 
\section{Related Works}
Camera trajectory generation has evolved significantly from traditional rule-based approaches \cite{he1996virtual,christie2008camera} to modern deep learning methods \cite{jiang2024cinematographic,courant2025exceptional}. Previous works in this domain can be broadly categorized into three main streams: rule-based methods that rely on predefined cinematographic principles \cite{silva2011automatic,galvane2014narrative}, optimization-based techniques that formulate trajectory generation as a constraint satisfaction problem \cite{bengio2000gradient,tewari2021overview}, and learning-based approaches that leverage data-driven models to generate or transfer camera movements \cite{jiang2020example, wang2023jaws}. In recent years, with the advancement of large language models and vision-language models like CLIP \cite{radford2021learning}, there has been growing interest in incorporating semantic understanding into camera control systems \cite{he2024cameractrl,liu2024chatcam}.

\textbf{Rule-based Approach: }Early work in camera trajectory generation primarily relied on rule-based methods that encoded established cinematographic principles into algorithmic frameworks. The Virtual Cinematographer \cite{he1996virtual} pioneered this approach by implementing film idioms as hierarchical finite state machines to automatically produce camera specifications in real-time. Building on this foundation, next work \cite{tomlinson2000expressive} developed a behavior-based system that utilized ethologically inspired mechanisms to adapt to interactive 3D environments. Upcoming work \cite{christie2008camera} further expanded these concepts by introducing a framework for viewpoint computation and motion planning based on established filming conventions. More recent work \cite{louarn2018automated} attempted to enable more flexible camera control by extending the Prose Storyboard Language (PSL) \cite{ronfard2015prose}. These rule-based approaches offered computational efficiency and reliability in controlled scenarios but were limited by their inherent rigidity, making them unsuitable for new, dynamic, or creative contexts.

\textbf{Optimization Approach:} Optimization-based methods emerged as a more flexible alternative to rule-based systems, formulating camera trajectory generation as a mathematical optimization problem. Early work computed optimal viewpoints by maximizing the projected area to surface area ratio \cite{kamada1988views}. The field advanced significantly with the introduction of Toric Space \cite{Lino2012}, which is a 2D manifold representation model that transformed complex camera control into a more efficient algebraic framework.  Later developments \cite{Lino2015} further refined this approach and reduced the search space to four degrees of freedom while maintaining precise control over visual composition. In the context of drone cinematography, several researchers \cite{nageli2017multi, bonatti2020autonomous} developed specialized optimization frameworks that balanced cinematographic constraints with physical drone limitations. However, the adaptation to new scenarios proved particularly challenging, as it necessitated reformulating the optimization problem and its constraints \cite{galvane2018directing}. Additionally, they often struggled with computational complexity in dynamic environments, particularly when dealing with multiple objectives or complex constraints \cite{dehghanian2025camera}.

\textbf{Deep Learning Approaches: }Recent years have witnessed a significant shift towards deep learning-based approaches for camera trajectory generation. Early explorations \cite{chen2016learning} utilized Recurrent Random Forests to predict camera pan angles in sports events, demonstrating the potential of learning-based methods for automated cinematography. Jiang et al. introduced an example-driven approach that could transfer camera behaviors from reference videos to new scenarios using a combination of LSTM networks and a mixture of experts framework \cite{jiang2020example}. They later extended the framework to incorporate precise frame control through keyframing and employ an autoregressive model for step-by-step camera position generation \cite{jiang2021camera}.

Diffusion models \cite{ho2020denoising} have made significant strides in the domain of camera trajectory generation, with several studies combining transformer architectures and diffusion-based models \cite{jiang2024cinematographic,li2024director3d,wang2024dancecamera3d}. The first work in this line, CCD \cite{jiang2024cinematographic}, was trained on synthetic datasets and laid the foundation for using diffusion models in this area. The E.T. framework \cite{courant2025exceptional} introduced a dedicated dataset along with three diffusion-based architectures that varied in their approach to incorporating conditional information. Among these, the architecture that integrated transformers to encode inputs, achieved superior performance. DCM model \cite{wang2024dancecamera3d} focuses on filming dance performances and integrates music and choreography to guide camera movements. However, DCM does not support conditioning through prompts, keyframes, or style constraints, limiting its flexibility compared to other models in this domain.

\textbf{textual Guidance:} The integration of Large Language Models (LLMs) and Vision-Language Models (VLMs) represents the latest frontier in camera trajectory generation \cite{he2024cameractrl,liu2024chatcam}. CLIP \cite{radford2021learning} has emerged as a particularly powerful tool for bridging textual descriptions with visual understanding, enabling more intuitive control over camera movements. CameraCtrl \cite{he2024cameractrl} demonstrated the potential of integrating CLIP embeddings with video diffusion models, enabling semantically aware camera control. The emergence of more sophisticated LLMs \cite{dubey2024llama, hurst2024gpt} has further enhanced the field's capability to interpret complex cinematographic instructions. These models excel at understanding high-level creative intent and translating it into concrete camera parameters. However, the challenge remains in balancing the semantic understanding provided by these models with the technical constraints of camera movement \cite{dehghanian2025camera}. 

\section{LensCraft}
\label{sec: model}
One of the key challenges that hindered previous approaches from achieving robust and generalizable results was the lack of a comprehensive and balanced training dataset.
Previous datasets for camera trajectory generation suffer from several critical limitations \cite{dehghanian2025camera}. First, they lack diversity in movement and shot types (e.g., RealEstate10K dataset \cite{zhou2018stereo, wang2024motionctrl} predominantly feature slow, smooth movements) or exhibit severe imbalances (e.g., 70\% of the E.T. dataset \cite{courant2025exceptional} consists of static shots), which can significantly bias model training. Second, they lack precise cinematographic annotations and labels that are essential for learning professional camera work \cite{jiang2024cinematographic, jiang2020example}. Third, their reliance on approximate SLAM algorithms \cite{goel2023humans} for camera and character position estimation introduces inherent inaccuracies that propagate through the training process. \cite{wang2024dancecamera3d}

To address these limitations, we present a comprehensive methodology for generating a high-quality, cinematography-focused dataset. Our approach begins with a user prompt which is then translates into specialized cinematographic language that enables the generation of high-level standardized descriptions. These descriptions are then transformed through our compiler system into precise, low-level simulation instructions. A custom simulator processes these representations to generate specific camera trajectories, resulting in clean, paired data containing high-level prompts, low-level representations, and their corresponding camera trajectories. The details of this stage are examined in Appendix 
\ref{app: dataset}.

\subsection{Architecture}
Building upon this foundation, we introduce LensCraft, a novel volume-aware camera trajectory generation model. Our model employs an innovative training strategy that enables simultaneous conditioning on multiple inputs: textual prompt, character trajectory and volume, keyframes, and source trajectories. This multi-modal conditioning allows LensCraft to generate camera trajectories that are both technically precise and creatively aligned with user intent. 
To harness this flexibility and context-awareness we decide to use a deep data-driven approach which is built on a Denoising Masked Autoencoder Transformer architecture. The core of LensCraft's design lies in its ability to learn and align representations across diverse modalities. To ensure precise conditioning of all inputs, we use a different strategy in training and inference stages. 

To make the model aware of the subject, we recognized that one of the most critical factors when filming is the volume of the subject. To effectively model this, we chose to use a bounding box (VBox) as a representation of the subject's volume. This representation allows us to easily capture the subject's volume during simulation and pass it to our model throughout the learning phase, ensuring accurate conditioning in the generated camera trajectories.

In the training phase, LensCraft Encoder takes as input the trajectory of the subject, VBox and the corresponding camera trajectory and tries to reconstruct the camera trajectory in an autoencoder structure. To guide the learning process, we utilize CLIP \cite{radford2021learning} as a teacher and force the model to align its middle embedding with the representations produced by CLIP. Inspired by the principles of multi-task learning \cite{caruana1997multitask}, we design the encoder to simultaneously learn and extract both high-level and low-level embeddings. Details of generating these embeddings can be found at Appendix \ref{app: dataset}.

Given that we generate the dataset and have access to both high-level and low-level embeddings of the prompts, we leverage the multi-task learning paradigm during the training of the encoder to strengthen it by encoding both high-level and low-level embeddings together. However, during the decoding stage, we pass only the high-level embeddings to the decoder.

The Decoder component of LensCraft takes solely high-level representations as input, in contrast to traditional decoder architectures that often rely on all of output encoder or skip connections. This design choice is motivated by our goal to ensure the model relies solely on these high-level cinematographic specifications to generate trajectories that align with professional standards while maintaining creative intent, rather than being overly constrained by low-level technical parameters.


As previously mentioned, the input to our architecture is based on a denoising masked autoencoder. The denoising aspect of our architecture serves a clear purpose: ensuring robustness and generating smooth trajectories by learning to recover clean signals from noisy inputs. However, our choice to incorporate masking goes beyond traditional noise handling - it provides a flexible mechanism for conditioning the model's output during inference. 

By strategically applying masks during training, we teach the model to handle partial information scenarios that mirror real-world use cases where frames may be constrained or known. We implement a progressive masking scheduler that allows the model to first learn basic trajectory patterns before adapting to increasingly sparse inputs. This masking strategy enables the model to seamlessly integrate a wide range of frame-based conditions during inference, from a few key frames to the full source trajectory, without requiring separate specialized architectures for each scenario. 

To support this progressive conditioning, we employ a continuous teacher forcing \cite{williams1989learning} strategy during training, which guides the decoder during training by interpolating between the encoder’s embeddings and ground truth CLIP embeddings.Here, we employ a weighted fusion where the decoder input is a convex combination of the encoder’s output and the CLIP embedding.  Early in training, the decoder relies more on the encoder’s output to encourage the model to learn robust autoencoding and reconstruction capabilities. . As training advances, the balance shifts toward CLIP embeddings—guiding the model to operate under inference-like conditions where only high-level semantic cues are available. 

\begin{figure}[t]
    \centering
    \begin{subfigure}[t]{0.49\linewidth}
        \centering
        \includegraphics[width=\linewidth]{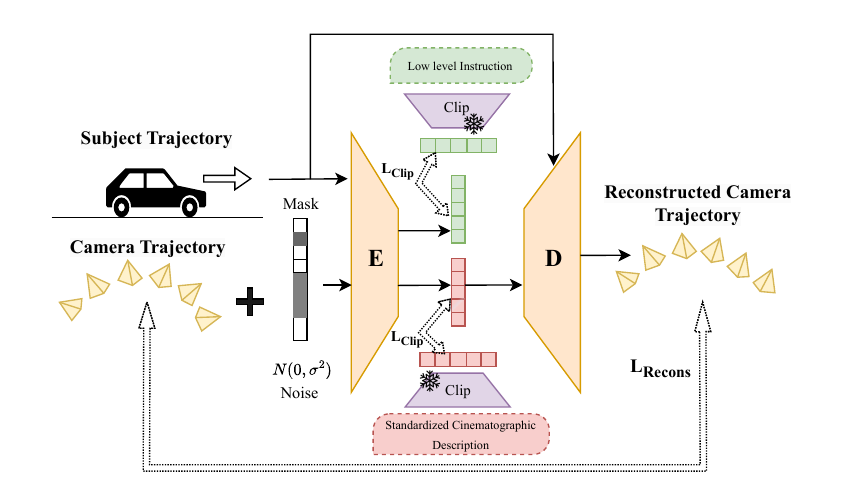}
        \caption{Training stage: multi-loss supervision and teacher forcing schedule.}
        \label{fig:sub_train}
    \end{subfigure}
    \hfill
    \begin{subfigure}[t]{0.49\linewidth}
        \centering
        \includegraphics[width=\linewidth]{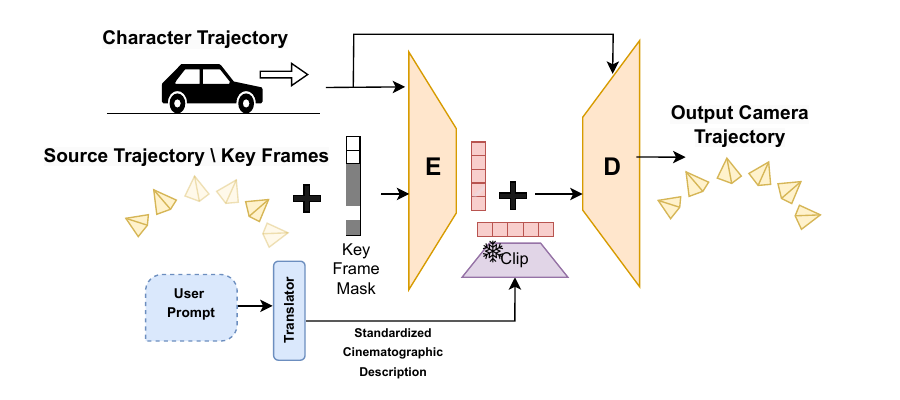}
        \caption{Inference stage: flexible conditioning using prompts, keyframes, and source trajectories.}
        \label{fig:sub_infer}
    \end{subfigure}
    \caption{End-to-end pipeline of the training (a) and inference (b) of LensCraft.}
    \label{fig:overview}
\end{figure}

By the end of training in this framework our model, LensCraft, learns to extract fine-grained spatial and temporal details from the encoder—when available—and integrate them with the global semantic intent provided by the textual prompt, resulting in precise aligned trajectory generation.
The overall pipeline of LensCraft, encompassing both the training (left) and inference (right) stages, is illustrated in Figure~\ref{fig:overview}, highlighting the flow of inputs, loss functions, and modality-specific conditioning mechanisms across the two phases.

The further detail of loss function used in training procedure is discussed in next subsection.





\subsection{Discrepancy Function}
In this work, we represent camera poses using a standard 6 Degrees of Freedom (6DoF) parameterization, which comprises three transitional components $(x, y, z)$ and three rotational components (Euler angles: $\phi, \psi, \chi$). It is critical to recognize that the transition and rotation components describe fundamentally different physical quantities and thus require distinct treatment when computing differences in the loss function design. 

To unify these aspects in a single differentiable metric, we define a composite discrepancy function $\bigl\| \cdot , \cdot \bigr\|$ that measures the distance between two camera poses and returns a scalar representing the overall difference. For the transition component, we utilize the Mean Squared Error (MSE) between the positional vectors, which accurately captures Euclidean distance between camera locations.

In cinematographic applications, even a small angular deviation can have a substantial impact on the output, potentially causing the subject to be missed entirely and rendering the shot meaningless.
So, given this level of importance, these two domains cannot be aligned simply by a constant scalar factor. Motivated by this, we devised an alternative definition for angular distance that more naturally reflects perceptual and framing errors.

 To correct angular errors so that the subject remains properly framed in the view, we map the angular discrepancy through a tangent function, effectively transferring this error into the Euclidean domain, and it shows how much would the camera's position need to change to observe the subject from the generated rotation angle instead of the ground truth? This geometric interpretation naturally relates angular differences to positional displacements, allowing rotations to be measured in the same modality and domain as translations.
 
 Formally, we define the discrepancy between two camera poses \( C = (x, y, z, \phi, \psi, \chi) \) and \( \hat{C} = (\hat{x}, \hat{y}, \hat{z}, \hat{\phi}, \hat{\psi}, \hat{\chi}) \) as a combined metric shown in Equation \ref{eq: distance}, simultaneously captures positional and angular difference within a single scalar value.
 
\begin{equation}
\label{eq: distance}
\bigl\|(C, \hat{C}) \bigr\| = 
\sqrt{
\underbrace{\sum_{i=1}^{3} (C_i - \hat{C}_i)^2}_{\text{MSE on transition}}
} +
\underbrace{
\sum_{j=4}^{6} \tan\left( \frac{\pi}{4 + \epsilon} + 1 - \langle n(\hat{\theta}_j), n(\theta_j) \rangle \right)
}_{\text{Angular distance on rotation}}.
\end{equation}

where \( n(\cdot) \) denote the normalized rotation matrix representation, \(\langle \cdot, \cdot \rangle\) represents the inner product, and \(\epsilon\) is a hyperparameter that controls the scale and relative importance of the angular component, while also ensuring numerical stability within the tangent function’s domain.

This distance metric thus integrates the Mean Squared Error on translation components with a geometrically meaningful, nonlinear mapping of angular discrepancies, yielding a stable and perceptually consistent objective for training our model.

\subsection{Loss Function}

Having established a precise definition for measuring the discrepancy between two camera poses, we can now formulate the loss function to train our model. To effectively capture the multifaceted nature of cinematographic trajectory generation—balancing low-level motion realism with high-level semantic intent—we design a composite loss function composed of five distinct terms. Each term addresses a specific aspect of the problem: (1) initial loss, which penalizes deviations at the initial point of the trajectory; (2) relational loss, ensuring consistency of relative spatial positions throughout the sequence; (3) speed loss, encouraging smooth temporal transitions and motion dynamics; (4)  CLIP loss, to semantically anchor the model’s internal representations to cinematographic language, and (5) cycle consistency loss, promoting preservation of semantic meaning through the encoding-decoding process.

 The initial loss defined as Equation~\ref{eq:init_loss}: 

\begin{equation}
\label{eq:init_loss}
L_{\text{init}} \;=\; \bigl\| C[0] , \hat{C}[0] \bigr\|
\end{equation}

This term explicitly penalizes initialization error at the beginning of the sequence, which is critical for preserving the initial framing and spatial context of the shot. Relative Loss is formulated as Equation \ref{eq:rel_loss}

\begin{equation}
\label{eq:rel_loss}
L_{\text{rel}} \;=\; \sum_{i} \Bigl\| (C[i]-C[0]) , (\hat{C}[i]-\hat{C}[0]) \Bigr\|
\end{equation}

It measures every frame’s displacement relative to the first frame and forces those relative offsets to match the ground truth. By anchoring all poses to a common reference, it corrects for global drift and guarantees that the predicted trajectory occupies the right region of space—regardless of how well consecutive steps line up—so the overall shape of the path remains faithful to the intended cinematic motion. Next loss for training our model is Speed Loss (Equation \ref{eq:speed_loss}): 

\begin{equation}
\label{eq:speed_loss}
L_{\text{speed}} \;=\;\sum_{i}\
      \Bigl\| (C[i]-C[i-1]) , (\hat{C}[i]-\hat{C}[i-1]) \Bigr\|.
\end{equation}
Tis loss enforces first‑order temporal consistency by aligning the inter‑frame motion deltas, thereby discouraging abrupt or erratic transitions across time. Together, the hyperparameters $\lambda_1$ and $\lambda_2$ govern the trade‑off between static alignment, global shape fidelity, and dynamic smoothness. 

For the CLIP loss, we can formulate it as Equation~\ref{eq:clip_loss}:
\begin{equation}
\label{eq:clip_loss}
L_{\text{clip}}=\sum_{k \in \{\text{high}, \text{low}\}}\Bigl(1 - \frac{\langle E_k , CLIP_k \rangle}
{\|E_k\|_{2}\;\|CLIP_k\|_{2}}\Bigr)
\end{equation}
where $E_k$ represents the encoder's output embeddings, $CLIP_k$ shows the CLIP embeddings for prompts, \(\langle \cdot, \cdot \rangle\) represents the inner product, and $\|\cdot\|_{2}$ denotes norm 2 of the embedding. By minimizing this loss, the encoder is trained to interpret and embed cinematographic instructions in a semantically grounded space, consistent with CLIP’s visual-language alignment. This enables the encoder to serve as a robust conditioning module during inference, effectively mixing prompt-driven high-level intent with structural details from auxiliary inputs such as keyframes or source trajectories. 

The cycle consistency loss, which has been widely used in image generation domain \cite{zhu2017unpaired, li2017alice,dehghanian2024anomaly}, is adapted to ensure the preservation of high-level cinematographic semantics through the encoding-decoding process. This loss is computed only on high-level features as shown in Equation~\ref{eq:cycle_loss}:
\begin{equation}
\label{eq:cycle_loss}
L_{\text{cycle}}=\sum_{k \in {\text{high}}}
\Bigl(1 - \frac{\langle E_k , E_k^{\text{re}}\rangle}
{\|E_k\|\ \|E_k^{\text{re}}\|}\Bigr)
\end{equation}

where $E_k$ represents the original encoder features and $E_k^{\text{re}}$ denotes the re-encoded features obtained by passing the decoder's output again back through the encoder. This cosine similarity-based loss ensures that the semantic meaning of cinematographic concepts is preserved through the encoding-decoding process. Finally, the total loss is a weighted combination of all three components:
\begin{equation}
\label{eq:total_loss}
L_{\text{total}}=L_{\text{init}} +\alpha   L_{\text{rel}} + \beta L_{\text{speed}} +\gamma L_{\text{clip}} + \lambda L_{\text{cycle}}
\end{equation}
where $\alpha$, $\beta$, $\gamma$, and $\lambda$  are weighting coefficient hyperparameters that balance the contribution of each loss term.

\section{Experiments} 
In this section, we demonstrate the validity of our model through extensive experiments. The experimental details, including the experiment setup, evaluation metrics, and dataset information, are provided in Appendix \ref{app: experimental detail}.

\subsection{Results}
Comparative evaluations of LensCraft against the previous SOTA methods in the prompt-to-camera task, E.T. \cite{courant2025exceptional}, and CCD \cite{jiang2024cinematographic}, are presented. These results, analyzed on both the static and dynamic subsets, are detailed in Table \ref{tab: main-result}.


\begin{table}[htbp]
    \centering
    \caption{Comparison of SOTA Methods on Static and Dynamic Trajectories}
    \label{tab: main-result}
    \resizebox{0.6\textwidth}{!}{
    \begin{tabular}{@{}lccccccc@{}}
        \toprule
        \textbf{Set} & \textbf{Methods} & \textbf{FID $\downarrow$} & \textbf{P $\uparrow$} & \textbf{R $\uparrow$} & \textbf{D $\uparrow$} & \textbf{C $\uparrow$} & \textbf{CS $\uparrow$} \\
        \midrule
        \multirow{5}{*}{\rotatebox{90}{\textbf{Static}}} 
        & CCD          & 320         & \textbf{0.69} & 0.19 & \textbf{0.57} & \uline{0.15} & 88.68 \\
        & E.T.\ A      & 166         & 0.15          & 0.53 & 0.09          & 0.10          & 90.62 \\
        & E.T.\ B      & 172.5       & 0.14          & 0.54 & 0.08          & 0.09          & 90.56 \\
        & E.T.\ C      & \uline{161.37} & 0.17          & \uline{0.55} & 0.09          & 0.10          & \uline{90.68} \\
        \cdashline{2-8}
        & LensCraft    & \textbf{40.4} & \uline{0.53} & \textbf{0.70} & \uline{0.33} & \textbf{0.29} & \textbf{92.75} \\
        \midrule
        \multirow{5}{*}{\rotatebox{90}{\textbf{Dynamic}}}
        & CCD          & 307.5          & \textbf{0.64}   & 0.21            & \textbf{0.51}             & \uline{0.14}       & 88.75 \\
        & E.T.\ A      & 151.62       & 0.23 & 0.60          & 0.13 & 0.09 & 90.68 \\
        & E.T.\ B      & 158.62       & 0.20         & \uline{0.62}  & 0.10         & 0.07         & 90.68 \\
        & E.T.\ C      & \uline{151.5}        & 0.19         & 0.61          & 0.10         & 0.07         & \uline{90.75} \\
        \cdashline{2-8}
        & LensCraft    &\textbf{24.35} & \uline{0.54} & \textbf{0.73} & \uline{0.38} & \textbf{0.40} & \textbf{92.93} \\
        \bottomrule
    \end{tabular}}
\end{table}

Table \ref{tab: main-result} reveals that in general LensCraft outperforms both CCD and E.T. across all metrics and subsets. In the static subset, E.T. exhibits more performance degradation, which can be attributed to its overemphasis on subject movement rather than adherence to input instructions. Conversely, CCD's performance drops significantly in the dynamic subset, primarily due to its reliance on a subject-specific coordinate system. This approach struggles to accurately model real-world dynamics, unlike our global coordinate framework. LensCraft, by contrast, demonstrates robust adaptability to varying trajectory complexities, delivering stable and reliable results across all conditions. Due to page limitations, the results of running our model, E.T., and CCD on these datasets are provided in Appendix \ref{app: run other datasets} showing the robustness and generalizability of our model.

\subsection{Multimodal Conditioning}
To evaluate the capability of our model in handling inputs from diverse domains, we tested various combinations of input modalities. These combinations were fed into the model, and its outputs were compared against the ground truth trajectories. The results of this evaluation are presented in Table \ref{tab: multimodal}, highlighting the model's performance across different multimodal configurations.

\begin{table}[htbp]
    \caption{Performance Comparison Across Multimodal Input Configurations}
    \label{tab: multimodal}
    \centering
    \resizebox{0.7\textwidth}{!}{
    \begin{tabular}{@{}llcccccc@{}}
        \toprule
        \textbf{Set} & \textbf{Input(s)} & \textbf{FID $\downarrow$} & \textbf{P $\uparrow$} & \textbf{R $\uparrow$} & \textbf{D $\uparrow$} & \textbf{C $\uparrow$} & \textbf{CS $\uparrow$} \\
        \midrule
        \multirow{5}{*}{\rotatebox{90}{\textbf{Static}}} 
        & Prompt Only                & 40.40          & 0.53              & \textbf{0.70} & 0.33              & 0.29              & 92.75 \\
        & Key Frame                  & \uline{10.76}  & \uline{0.57}      & 0.60          & 0.37              & \uline{0.38}      & 92.81 \\
        & Source trajectory          & 12.79          & 0.54              & 0.65          & 0.33              & 0.34              & \uline{93.06} \\
        & Prompt + Key Frame         & \textbf{9.68}  & \textbf{0.59}     & 0.60          & \textbf{0.39}     & \textbf{0.41}     & 92.87 \\
        & Prompt + Source trajectory & 15.38          & \textbf{0.59}     & \uline{0.67}  & \uline{0.38}      & 0.37              & \textbf{93.18} \\
        \midrule
        \multirow{5}{*}{\rotatebox{90}{\textbf{Dynamic}}}
        & Prompt Only                & 24.35 & \uline{0.54} & \textbf{0.73} & \textbf{0.38} & \textbf{0.40} & 92.93 \\
        & Key Frame                  & 13.54          & 0.52              & 0.55          & 0.32              & \uline{0.34}     & 92.62 \\
        & Source trajectory          & \textbf{11.70} & 0.50              & 0.62          & 0.30              & 0.31      & \uline{93.00} \\
        & Prompt + Key Frame         & \uline{12.53}  & \uline{0.53}      & 0.59          & 0.33      & \uline{0.34}     & 92.43 \\
        & Prompt + Source trajectory & 13.70          & \textbf{0.56}     & \uline{0.64}  & \uline{0.35}     & \uline{0.34}     & \textbf{93.12} \\
        \bottomrule
    \end{tabular}}
\end{table}

As shown in Table \ref{tab: multimodal}, the model performs well under single-input conditioning, whether using prompts, keyframes, or source trajectory individually. However, the combined input configurations achieve superior results compared to any single modality. It should be mentioned that for testing key frame we randomly chose 1 to 10 key frames from test set, while in the source trajectory conditioning, we pass the full camera trajectory into the encoder; and the output reflects the model's reconstruction capability. In contrast, the prompt-only configuration demonstrates the model's generation capacity. The combined inputs highlight the model's ability to leverage both reconstruction and generation capabilities effectively. Additional qualitative results and visualization of multimodal conditioning can be found at Appendix \ref{app: qualitative_result}


\subsection{Ablation Study}

In this section, we analyze the effectiveness of various components of Loss and design choices in the architecture of LensCraft. We evaluate the model under different configurations to highlight the impact of these elements.  The results of the ablation study are presented in Table \ref{tab: ablation}.



\begin{table}[htbp]
    \caption{Ablation Study on Different Components of LensCraft}
    \label{tab: ablation}
    \centering
    \resizebox{0.6\textwidth}{!}{
        \begin{tabular}{@{}clcccccc@{}}
            \toprule
            \textbf{Variant} & \textbf{FID $\downarrow$} & \textbf{P $\uparrow$} & \textbf{R $\uparrow$} & \textbf{D $\uparrow$} & \textbf{C $\uparrow$} & \textbf{CS $\uparrow$} \\
            \midrule
            $L_{\text{clip}}$ \ding{55}            & 214.12 & 0.03 & \uline{0.71} & 0.01 & 0.03 & 0.00 \\
            $L_{\text{first frame}}$ \ding{55}     & 118.31 & 0.22 & 0.69          & 0.11 & 0.11 & 90.87 \\
            $L_{\text{relative}}$ \ding{55}        &  88.18 & 0.18 & 0.66          & 0.10 & 0.13 & \uline{92.75} \\
            $L_{\text{speed}}$ \ding{55}           &  
             34.34          & \uline{0.49}              & 0.66 & \uline{0.29}              & 0.27              & 92.25\\
            $L_{\text{cycle}}$ \ding{55}           &  28.78  & 0.36 & 0.70 & 0.28 & \textbf{0.41} & 92.56 \\
            Teacher \ding{55}                      & \uline{24.60} & 0.05 & 0.68 & 0.02 & 0.03 & 91.87 \\
            \cdashline{1-7}
            LensCraft                              & \textbf{24.35}  & \textbf{0.54} & \textbf{0.73} & \textbf{0.38} & \uline{0.40} & \textbf{92.93} \\
            \bottomrule
        \end{tabular}
    }
\end{table}

One notable strength of our model is its size and speed. Despite achieving state-of-the-art quality, our model operates with a smaller architecture compared to previous models. To further substantiate this claim, additional experiments on the speed of our model are explored in Appendix \ref{app:speed_experiment}.


\subsection{Embedding Visualization}
\begin{figure}[htbp]
    \centering
    \begin{minipage}[b]{0.32\textwidth}
        \centering
        \includegraphics[width=\textwidth]{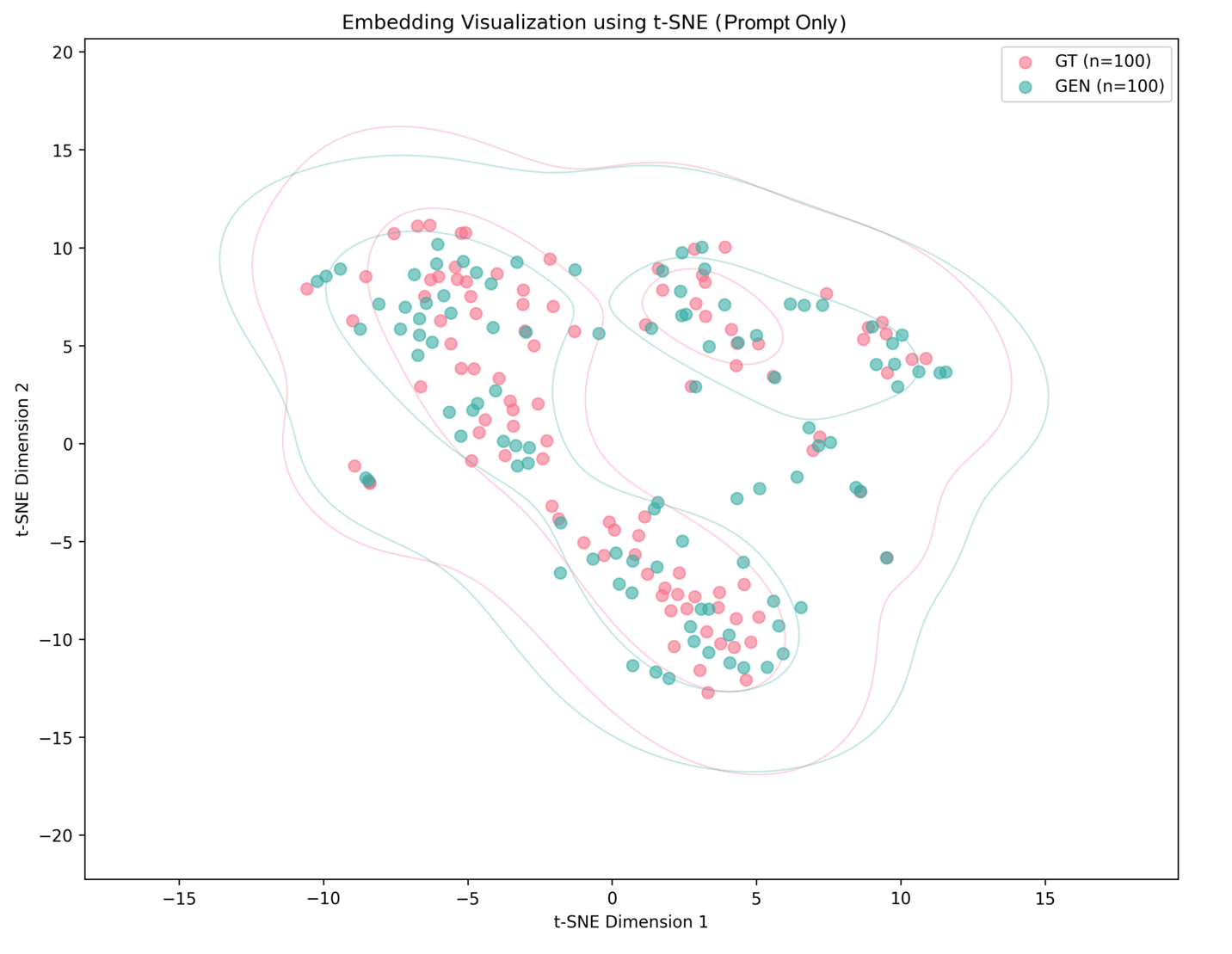}
        \caption*{(a) Prompt Only}
    \end{minipage}
    \hfill
    \begin{minipage}[b]{0.32\textwidth}
        \centering
        \includegraphics[width=\textwidth]{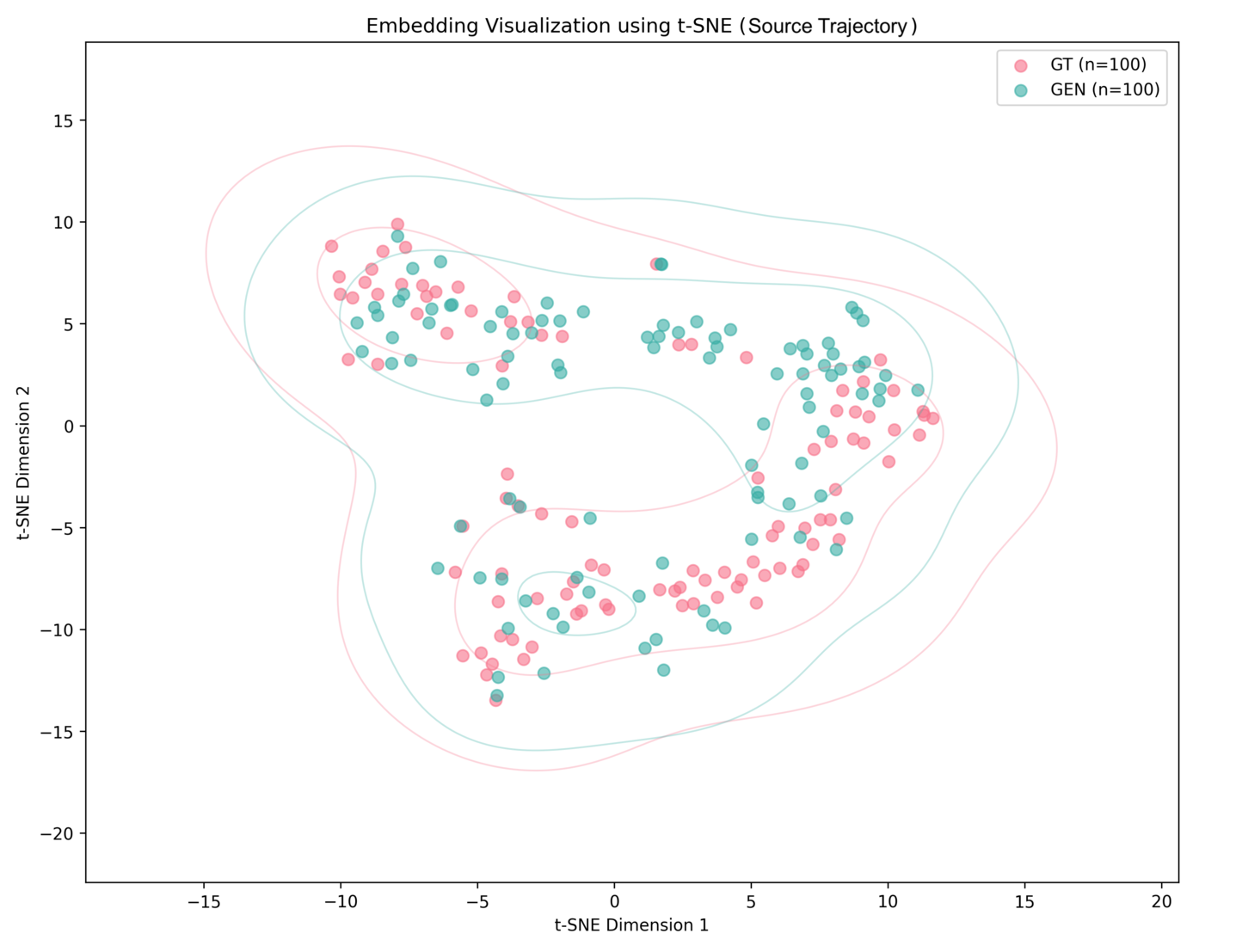}
        \caption*{(b) Source Trajectory}
    \end{minipage}
    \hfill
    \begin{minipage}[b]{0.32\textwidth}
        \centering
        \includegraphics[width=\textwidth]{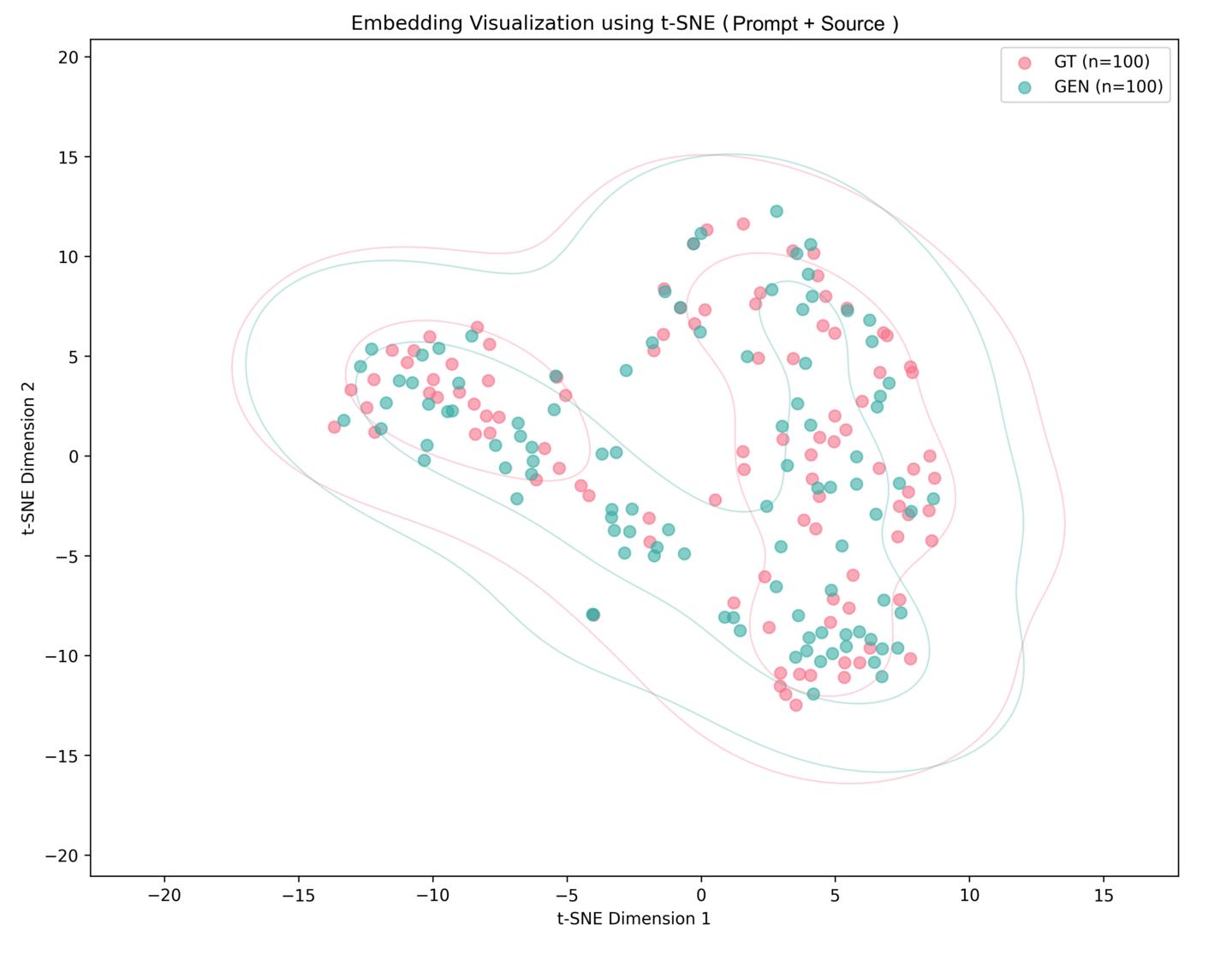}
        \caption*{(c) Prompt + Source Trajectory}
    \end{minipage}
    \caption{t-SNE visualizations of different conditioning setup.}
    \label{fig:tsne}
\end{figure}

embeddings of the input prompts alongside the embeddings extracted from the generated trajectories under different conditioning configurations. Specifically, after generating camera trajectories from various input modalities (e.g., prompt-only, keyframes, source trajectory), we re-encode these trajectories using our encoder to obtain their latent embeddings. This allows us to directly compare and analyze the semantic structure of these representations against the original CLIP embeddings of the prompts. As illustrated in Figure~\ref{fig:tsne}, we use t-SNE \cite{vanDerMaaten2008} for dimensionality reduction to project both sets of embeddings into a 2D space. 


As expected, the most coherent and well-aligned embedding clusters emerge when both the source trajectory and textual prompt are used jointly as input (c). This configuration provides the richest conditioning signal, allowing the model to generate trajectories that are semantically and structurally aligned with the intended cinematographic instruction. Notably, we also observe a significant overlap between the latent representations of the prompt-only (a) and source trajectory (b) generations. This intertwined structure in the embedding space suggests that the model effectively captures shared semantic cues across different modalities. It demonstrates the strong ability of our model to learn a unified latent space that generalizes well across varying input conditions, maintaining meaningful semantic continuity even when only partial information is available.



\subsection{Limitations}
Our evaluation demonstrates LensCraft's effectiveness in bridging automated camera control and professional cinematography through its unified multi-modal architecture. However, technical limitations exist in the current implementation. Like prior approaches \cite{jiang2024cinematographic,courant2025exceptional,wang2024dancecamera3d}, LensCraft only supports single-subject character, limiting its applicability in multi-subject scenarios. Also, while the introduction of the interest box provides a formal representation of directorial framing requirements, its fixed configuration limits adaptability to fine-grained shot specifications. (e.g., "close-up on character's hand"), potentially compromising precise desired subject framing. 


\section{Conclusion}
In this paper, we introduced LensCraft, a novel approach to automated camera trajectory generation that successfully connects professional cinematographic execution and precise user control. Our primary contributions include a comprehensive cinematography language and simulation framework that enables high-quality, balanced training data generation, and an innovative architecture leveraging multi-modal conditioning through CLIP embeddings and progressive masking strategies. Through extensive experimentation, we demonstrated that LensCraft consistently outperforms existing SOTA methods across all evaluation metrics, showing particular strength in maintaining professional quality while adhering to user intent and higher semantic alignment. Looking forward, LensCraft represents a significant step toward democratizing professional cinematography by providing an intuitive interface for specifying complex camera movements while maintaining high-quality output.



\bibliographystyle{ACM-Reference-Format}
\bibliography{sample-base}

\newpage
\appendix
\section{Dataset Generation}
\label{app: dataset}
\subsection{User Prompt Processing}
A fundamental obstacle in automated cinematography
is translating unstructured natural language input prompts into executable cinematographic parameters. Especially users those without professional filming backgrounds, tend to express their creative intent through informal, context-rich descriptions that may lack technical precision. For example, a user might request "make the camera swoop dramatically around the car while keeping it centered" rather than specifying exact angles and movements.

To address this challenge, we developed a  multi-stage input processing pipeline centered around a RoBERTa model \cite{liu2019roberta} as translator. The training data for this translator was generated through a novel permutation-based approach leveraging large language models. First, we systematically generated a comprehensive set of standardized cinematographic descriptions (SCD) by permuting all possible combinations of parameters (which the details will be described in next section). These formal descriptions were then transformed into natural language prompts using GPT-4.1 nano (a lightweight variant of GPT-4 \cite{openai2023gpt4}), which generated both complete and partial specifications mimicking real-world user inputs.

This synthetic data generation process allowed us to create a robust training set covering the full space of possible camera movements while ensuring accurate alignment between natural language prompts and their corresponding technical specifications. The translator can thus handle both informal, incomplete user prompts and precise technical descriptions during inference time. 

The translator employs attention mechanisms to identify relevant cinematographic elements within the prompt while maintaining contextual understanding. For instance, words like "swoop" and "dramatically" are mapped to specific movement patterns and timing parameters, while spatial relationships ("around", "centered") are translated into concrete positioning and framing requirements.Another important aspect of this translation process is handling ambiguity and incomplete specifications. When user prompts lack certain technical details, the system applies intelligent defaults based on cinematographic best practices. For example, if a user requests a circular movement without specifying the height, the system will maintain the current camera status unless there are contextual clues suggesting otherwise.

To assess the effectiveness of our translator, we conducted a controlled experiment using 1000 synthetic user prompts generated by GPT-4.1 nano. Each prompt was processed by our translator to produce a standardized cinematographic description, which we then evaluated through retrieval tasks. Specifically, we computed R\@k using CLIP embeddings to determine how often the correct cinematographic description appeared among the top-k nearest neighbors, following the standard retrieval evaluation protocols used in prior works \cite{guo2022generating, petrovich2023tmr}. A higher R\@k indicates better alignment between natural language input and standardized cinematographic intent. Table \ref{tab: translator} presents standard retrieval performance measures.

To assess the effectiveness of our translator, we conducted a controlled experiment using 1000 synthetic user prompts generated by GPT-4.1 nano. Each prompt was processed by our translator to produce a standardized cinematographic description. We evaluated the translator by selecting the top-$k$ most probable outputs predicted by the model, where the probability of each output class was derived from the model's output logits. We then checked whether the ground truth (GT) standardized description was among these top-$k$ predictions. This approach allows us to assess the model’s confidence-aligned ranking of outputs. Specifically, we computed Recall@$k$ (R@$k$) to measure how often the correct description appeared within the top-$k$ predictions, following the standard evaluation protocol used in prior works~\cite{guo2022generating, petrovich2023tmr}. A higher R@$k$ indicates better alignment between natural language input and standardized cinematographic intent. Table~\ref{tab:translator} presents the resulting recall performance.

\begin{table}[h]
\centering
\caption{Recall@k results for prompt-to-SCD retrieval task.}
\label{tab:translator}
\begin{tabular}{@{}lcccccccccc@{}}
\toprule
\textbf{R@1} $\uparrow$ & \textbf{R@2} $\uparrow$ & \textbf{R@3} $\uparrow$ & \textbf{R@4} $\uparrow$ & \textbf{R@5} $\uparrow$ & \textbf{R@6} $\uparrow$ & \textbf{R@7} $\uparrow$ & \textbf{R@8} $\uparrow$ & \textbf{R@9} $\uparrow$ & \textbf{R@10} $\uparrow$ \\
\midrule
67.65 & 78.58 & 84.87 & 89.65 & 92.82 & 94.70 & 95.82 & 96.78 & 97.63 & 97.93 \\
\bottomrule
\end{tabular}
\end{table}

\subsection{Standardized Cinematographic Description}
\label{sec: cinematographic description}
A key challenge in developing automated camera trajectory systems is designing a user interface that balances professional cinematographic needs with practical usability. Existing approaches tend toward two extremes: formal cinematographic languages \cite{louarn2020interactive, wu2018thinking} that are too complex and detailed for practical use, or oversimplified controls that fail to capture the requirements and nuances of professional camera work \cite{jiang2024cinematographic,courant2025exceptional}.

To achieve this balance, we conducted a series of exploratory consultations with a diverse range of users, including professional cinematographers, experienced film critics, and film school students. During these sessions, participants shared insights on how they might verbally direct camera movements to achieve specific visual outcomes and reviewed generated trajectories to provide feedback on alignment with their intent. Based on their input and multiple rounds of refinement, we identified four key parameters for describing our standardized description: the shot type (such as close-up or long shot), camera angle, character side, framing composition—defined for the beginning and optionally end of the shot—and the movement type and easing which is speed characterizing transitions between these states.

To achieve this balance, we conducted a series of iterative consultations with a diverse range of users, including professional cinematographers, experienced film critics, and film school students. Through multiple validation sessions, participants were asked to articulate how they would verbally direct camera movements to achieve specific visual outcomes, followed by reviewing and critiquing generated trajectories for alignment with their intent. After multiple rounds of refinement and validation, we distilled their requirements down to four fundamental parameters that together form a complete and minimal representation for describing high level user prompt: the shot type (such as close-up or long shot), camera angle, and framing composition - all defined for both the beginning and optionally end of the shot - along with the movement type and speed that characterizes how the camera transitions between these states throughout the shot. 


\subsection{Instruction Alignment}
From this standardized cinematographic description, our system generates simulation instructions that consist of four key components: detailed setup for initial and end points, movement constraints, and interpolation specifications. In cases where the user does not specify an end point, we randomly select a plausible combination that aligns with the initial setup and the desired movement characteristics.
The setup of both initial and end points follows a two-stage alignment process. 

In the macro stage, we first determine the optimal camera position in 3D space, treating the camera as a point that needs to be strategically placed to achieve the desired shot type and angle. This is followed by a micro alignment stage, where we fine-tune specific camera parameters such as orientation, field of view, and focus to achieve precise framing and composition.

In the macro alignment stage, we directly leverage two of our high-level parameters - shot type and camera angle - to determine the camera's initial position in 3D space. More formally, these parameters map directly to spherical coordinates \cite{Harvard1989} around the subject: The camera angle parameters map to the spherical angles - the vertical angle determines the polar angle ($\phi$), while the horizontal angle sets the azimuthal angle ($\theta$).

To model shot types effectively, we introduce a dual-scale system with two bounding boxes: the attention-bounding box (ABox) focused on the subject's key features, and the volume-bounding box (VBox) encompassing the entire subject. This dual-box representation is motivated by cinematographic principles - for example, in a close-up of a flower, the focus should be on its petals, while in a human subject, the camera should focus on the face.

To implement this, we compute a region of interest (ROI) that serves as the subject's bounding box for camera framing. For shots ranging from close-up to full shot, the ROI is computed through interpolation between ABox and VBox. To apply this interpolation, we use two parameters to compute the final ROI: an interpolation factor and a scale factor.
\begin{itemize}

 \item The interpolation factor: determines the blend between ABox and VBox. A value of 0.0 uses only the ABox (tight focus on key features), while a value of 1.0 uses only the VBox (full subject volume). Intermediate values linearly interpolate between the two, producing a box that balances detail and context based on the shot type.

\item The scale: controls how much of the interpolated box is ultimately shown in the frame. This simulates the level of zoom or framing distance.
    \item 
\end{itemize}

Together, these parameters allow for precise, shot-type-driven control of the ROI, supporting a wide range of cinematographic compositions—from tight, emotional portraits to sweeping, contextual vistas, the details of them is shown in Table \ref{tab: shot-scales}:
\begin{table}[h]
\centering
\caption{Shot Type Scaling Factors}
\begin{tabular}{llcc}
\toprule
\textbf{Abbrev.} & \textbf{Shot Type} & \textbf{Factor} & \textbf{Scale}  \\
\midrule
ECU & Extreme Close-Up & 0.0 & 0.5  \\
CU & Close-Up & 0.0 & 1.0  \\
MCU & Medium Close-Up & 0.25 & 1.0  \\
MS & Medium Shot & 0.50 & 1.0 \\
FS & Full Shot & 1.0 & 1.0  \\
LS & Long Shot & 1.0 & 1.5  \\
VLS & Very Long Shot & 1.0 & 2.0  \\
ELS & Extreme Long Shot & 1.0 & 3.0 \\
\bottomrule
\end{tabular}
\label{tab: shot-scales}
\end{table}

For instance, in an \textit{Extreme Close-Up (ECU)}, the camera focuses tightly on a small detail---such as a character’s eye---so we use only the ABox (interpolation factor $= 0.0$) and apply a scale of $0.5\times$, showing just half of the ABox for the most intimate framing. On the other end, an \textit{Extreme Long Shot (ELS)} requires maximum environmental context. In this case, we use only the VBox (interpolation factor $= 1.0$) and scale it by $3.0\times$, effectively pulling the camera back to emphasize setting and atmosphere.

\begin{figure}
\centering
\includegraphics[width=0.7\linewidth]{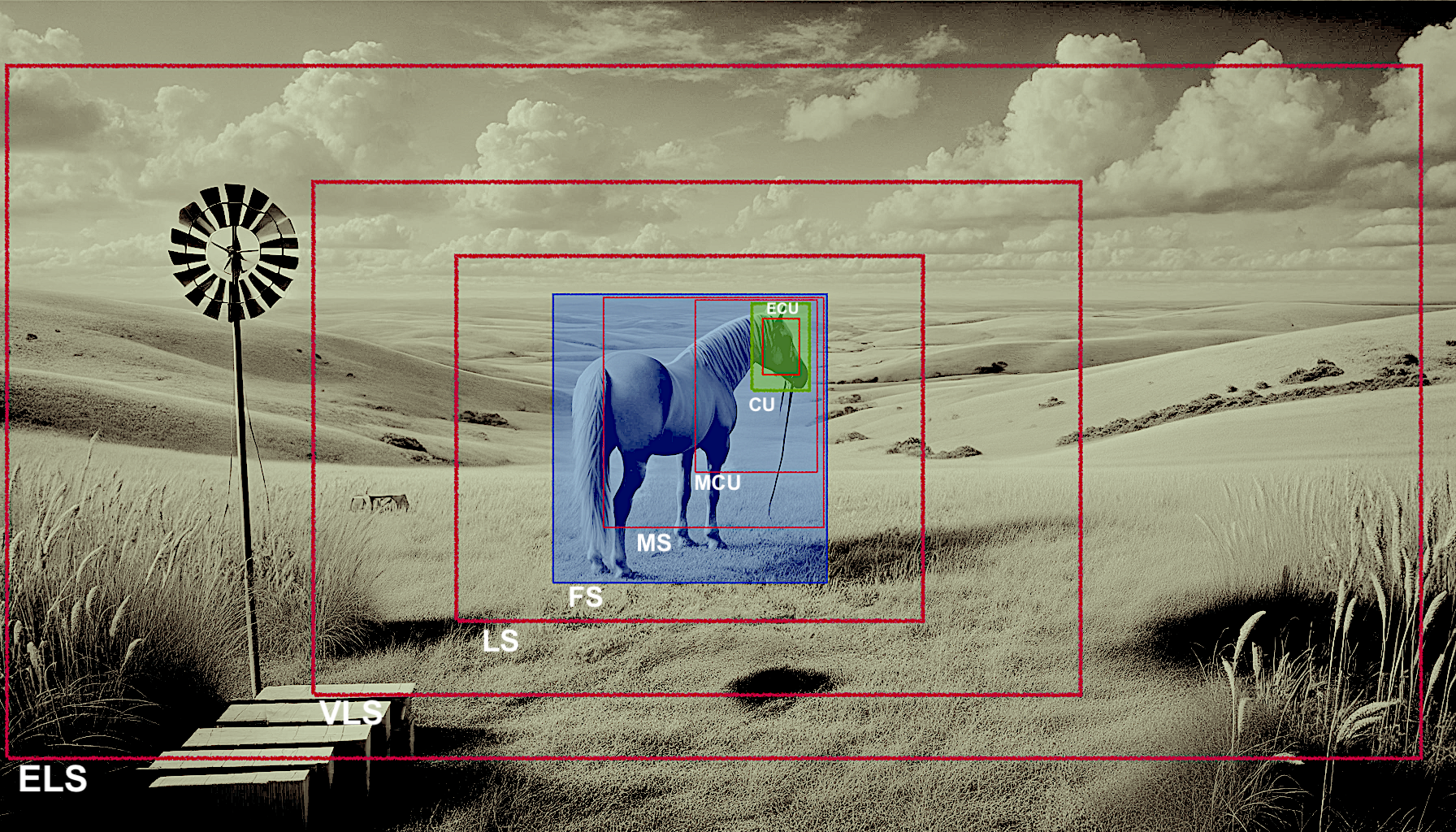}
\caption{Shot size modeling using dual bounding boxes. The blue overlay shows the progressive transition between ABox focusing on subject's key features, and VBox encompassing the entire subject.}
\label{fig: shot-size}
\end{figure}

The result of applying these rules on one sample is illustrated in Figure \ref{fig: shot-size}, from the most expansive establishing shot (ELS) to the most intimate detail shot (ECU). This two-step process allows for precise control over both the focus area and the amount of context shown in the frame. 

To model camera movement in our low-level representation, we use two key components: interpolation type and constraints. The interpolation can be either linear or subject-aware, representing different ways of transitioning between initial and final camera setups. For a camera path $P(t)$ where $t \in [0,1]$, given initial camera setup $P_0$ and final setup $P_1$, the two interpolation types are defined as follows:

Linear interpolation simply connects the start and end points:
$$P_{linear}(t) = (1-t)P_0 + tP_1$$

Subject-aware interpolation considers the character position $C$ and creates a curved path that maintains awareness of the subject:
$$P_{aware}(t) = P_0 + t(P_1 - P_0) + \alpha(1-t)t(\vec{d_0} - \vec{d_1})$$

where $\vec{d_0} = C - P_0$ and $\vec{d_1} = C - P_1$ represent the directional vectors from the camera to the subject at the start and end positions, respectively, and $\alpha$ is a scaling factor that controls the curve's magnitude. This formulation ensures the camera path naturally arcs around the subject rather than moving in a straight line.

For movement constraints, our consultations with cinematographers led us to define four fundamental constraints that effectively model different categories of camera movement:
\begin{enumerate}
    \item Static Location: This constraint locks the camera's position in space, allowing adjustments only through orientation and focal length changes. This is particularly useful for movements like panning where the camera pivots from a fixed position.
    \item Static Distance: This ensures the camera maintains a constant distance from the subject throughout the shot, making it ideal for orbiting movements where the camera circles the subject.
    \item Visibility at all: This constraint ensures continuous visual contact with the subject throughout the movement, preventing any temporary loss of sight during complex camera paths.
    \item Maximum Acceleration ($a_{max}$): This defines the upper limit of camera acceleration along the path, ensuring that generated movements remain physically feasible and can be executed smoothly in real-world scenarios.
\end{enumerate}

In the micro-alignment stage, we refine the camera setup through precise adjustments to focal length and orientation to achieve the desired framing composition. To handle framing specifications, we divide the camera's view plane into a nine-section grid following the rule of thirds \cite{malevs2012compositional,amirshahi2014evaluating}. Based on the user's framing requirements, we first attempt to achieve the desired subject placement through minimal adjustments to camera orientation. When orientation adjustments are constrained or insufficient, we make small modifications to the camera's position while maintaining the established shot type and angle. The focal length is then fine-tuned to ensure the subject appears at the correct scale within the specified grid section, complementing the distance-based scaling from the macro stage.

To validate the precision and quality of our low-level representation, we implemented an open-source, web-based simulation framework that generates the outcome camera trajectory according to the low-level instructions. This simulator which is is publicly available at \href{https://camera-simulation-virid.vercel.app/}{link},serves as both a development tool and a validation platform. The framework enables real-time visualization of camera trajectories and  provide a transparent and reproducible foundation for future research in this domain.

\section{Experimental Details}
\label{app: experimental detail}
\subsection{Implementation Details}
We implemented LensCraft in PyTorch and trained it on an NVIDIA GeForce RTX 3090 Ti GPU for 100 epochs ($\sim$6 hours). The model uses 4 transformer layers, 8 attention heads, 2048 MLP dimension, and 512 hidden size (matching CLIP-base). Training employs AdamW optimizer (lr=1e-4, $\beta_1$=0.9, $\beta_2$=0.95, weight decay=0.1) with batch size 128 Loss weights are set to uptuna results as $\alpha$=8.0, $\beta$=20, $\gamma$=50, $\lambda = 5$, with teacher forcing ratio increasing from 0.7 to 1.0. We implement progressive training with masking ratio increasing from 0.1 to 0.8 and noise ratio decreasing from 1.0 to 0.0. For sequences, we use 30-frame temporal windows with sampling for longer inputs. Camera trajectories use standard 6-DOF representation \cite{christie2008camera}, while for character trajectories, we modeled the 3D center position with its dimensions (width, length, height) and three directional parameters to capture temporal and spatial dynamics effectively.

\subsection{Metrics}
We evaluate our model using a comprehensive set of quantitative metrics that assess both the quality of generated trajectories and their semantic alignment with input descriptions. For distribution-level evaluation, we employ Fr\'{e}chet Inception Distance (FID) \cite{heusel2017gans} to measure the statistical similarity between real and generated trajectory distributions in feature space:

\begin{equation}
    \text{FID}(\mathcal{P}_r, \mathcal{P}_g) = \|\mu_r - \mu_g\|^2_2 + \text{Tr}(\Sigma_r + \Sigma_g - 2\sqrt{\Sigma_r\Sigma_g})
\end{equation}

\noindent where $\mu_r$, $\mu_g$ represent means and $\Sigma_r$, $\Sigma_g$ denote covariance matrices of real and generated distributions respectively.

To provide a more nuanced assessment of generation quality, we utilize four complementary manifold-based metrics redefined for generative models \cite{naeem2020reliable}: 
 Precision (P), Recall (R), Density (D), and Coverage (C). Precision quantifies the proportion of generated samples that align with the real data manifold:

\begin{equation}
    \text{P} = \frac{1}{M}\sum_{j=1}^M \mathds{1}_{Y_j \in \text{manifold}(X_1,...,X_N)}
\end{equation}

In this formulation, $M$ denotes the total number of generated samples,$N$ is the total number of real (ground-truth) samples. The indicator function 
$\mathds{1}_{z \in S}$ to denote membership of an element $z$ in a set $S$, where it evaluates to 1 if $z \in S$ and 0 otherwise. Also, $\text{manifold}( \dots)$ defines as the union of $k$-nearest neighbor (k-NN) balls centered at each sample of it. Next metric, Recall measures the proportion of real data represented in generated samples:

\begin{equation}
    \text{R} = \frac{1}{N}\sum_{i=1}^N \mathds{1}_{X_i \in \text{manifold}(Y_1,...,Y_M)}
\end{equation}

Density provides a more nuanced view of sample distribution within the manifold:

\begin{equation}
    \text{D} = \frac{1}{kM}\sum_{j=1}^M\sum_{i=1}^N \mathds{1}_{Y_j \in B(X_i,\text{NND}_k(X_i))}
\end{equation}

In the Density formulation, $k$ is the number of nearest neighbors considered when constructing the neighborhood around each real sample $X_i$. The function $\text{NND}_k(X_i)$ denotes the distance from $X_i$ to its $k$-th nearest neighbor among the real samples. The notation $B(X_i, \text{NND}_k(X_i))$ represents a  neighborhood of radius $\text{NND}_k(X_i)$ centered at $X_i$. To assess the diversity of generated samples, we use Convergence metric:

\begin{equation}
    \text{C} = \frac{1}{N}\sum_{i=1}^N \mathds{1}_{\exists j \text{ such that } Y_j \in B(X_i,\text{NND}_k(X_i))}
\end{equation}
The Coverage metric is conceptually related to Diversity, but instead of counting the number of neighboring generated samples, it evaluates whether at least one generated sample exists within the local neighborhood of each real sample. It indicates how well the generated distribution spans the support of the real data manifold.

Additionally, for evaluating semantic alignment of our trajectories, we compute CLIP-Score (CS) \cite{radford2021learning} which measures the cosine similarity between the embeddings of trajectories and their corresponding ground truth textual prompts in the CLIP feature space as:

\begin{equation}
    \text{CLIP-S}(c, v) = \sum_{i=1}^{N} \frac{c_i \cdot v_i}{\|c_i\|_2 \, \|v_i\|_2}
\end{equation}

where $c_i$ and $v_i$ denote the $i$-th Clip embeddings of the generated camera trajectory and the real textual prompt respectively, and $\| \cdot \|_2$ denotes the L2 norm.

\subsection{Dataset}
\label{sec: dataset}
We constructed the dataset comprising camera trajectories and paired textual descriptions, generated via the simulation Described in Appendix \ref{app: dataset}. For our experiments, the dataset was divided into two subsets. The static subset includes samples with stationary character motion trajectories, designed to focus on evaluating the model's cinematography understanding in a simplified context. Conversely, the dynamic subset involves moving subjects, challenging the model to account for subject dynamics and emphasizing its attentional capabilities.The dataset contains 100,000 samples, representing approximately 3 million frames, and is evenly split between static and dynamic subject trajectories (50
Additionally, to test the generalizability and adaptability of our model in real-world settings, we also evaluate it on the E.T. \cite{courant2025exceptional} and CCD \cite{jiang2024cinematographic} datasets.
Since the CCD and E.T. datasets represent subjects as points and do not provide height or volumetric data, we estimate the subject's volume using an average human height of 170 cm to construct a corresponding VBox representation. Additionally, we apply frame sampling to these datasets to align their temporal resolution and sequence length with our own, ensuring that our model can process them using its standard input frame size.

\section{Results on Other Datasets}
\label{app: run other datasets}
To evaluate the generalizability of our model beyond our own dataset, we test LensCraft on E.T. dataset \cite{courant2025exceptional}. We use the original textual prompt as input to all three models. As described in our inference stage, we process each prompt through our translator module to convert it into our standardized cinematographic language (SCL) before feeding it into our decoder module to generate the final trajectory.

\begin{figure}[t]
  \centering
  \includegraphics[width=\textwidth]{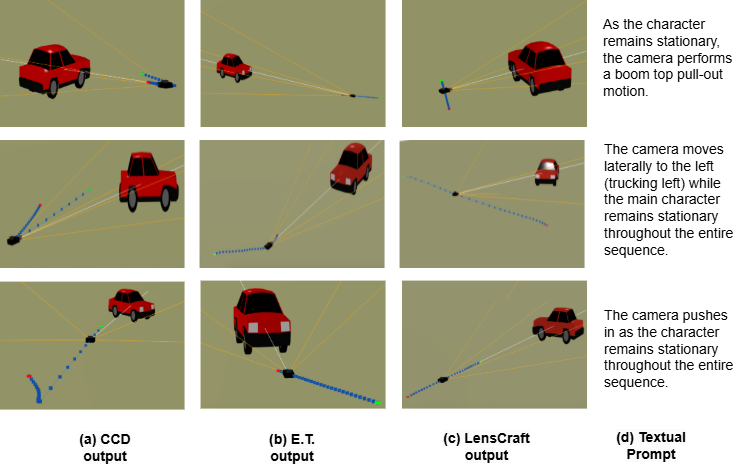}
  \caption{Comparison of model outputs on the E.T. dataset.}
  \label{fig:et_data}
\end{figure}

To evaluate the generalizability of our model beyond our own dataset, we test LensCraft on the E.T. dataset \cite{courant2025exceptional}. We use the original textual prompt as input to all three models. As described in our inference stage, we process each prompt through our translator module to convert it into our standardized cinematographic language (SCL) before feeding it into our decoder module to generate the final trajectory. To ensure that the evaluation focuses purely on camera behavior and is not influenced by subject motion complexity, we restrict our analysis to the static subject setting, which simplifies the scene and allows for a more direct comparison of camera control between different models. Figure~\ref{fig:et_data} visualizes the outputs of all models.

From right, the first row presents the original textual prompt from the E.T. dataset, and the next three rows show the trajectories generated by our model (LensCraft), the E.T. model, and the CCD model. While there exists a domain shift between our training data and the data distribution of this external dataset (both in terms of subject representation and cinematographic style) our model demonstrates robust generalization. LensCraft consistently produces smoother and more coherent camera motions, with trajectories that remain semantically faithful to the prompts and stylistically aligned with professional cinematographic conventions, outperforming prior models.

\section{Speed Comparison}
\label{app:speed_experiment}
To further validate the efficiency of our proposed model, we conduct a comparative study of inference speed against state-of-the-art models—CCD and E.T. This analysis includes two key aspects of computational cost: inference time and floating-point operations per second (FLOPs). To ensure a fair comparison, we adjusted the input frame size of LensCraft to match the temporal resolution used by E.T. and CCD. We also use the same batch size (128) for all models. These results are summarized in Table \ref{tab: speed}.

\begin{table}
\centering
\caption{Comparison of computational efficiency across models.}
\label{tab: speed}
\begin{tabular}{lcc}
\toprule
\textbf{Model} & \textbf{Inference Time (s) $\downarrow$} & \textbf{FLOPs (G) $\downarrow$}  \\
\midrule
CCD & 72.49 $\pm$ 3.87 & 243.44 \\
E.T. A& 43.21 $\pm$ 1.02 & 37.17 \\
E.T. B& 43.9362 $\pm$ 0.99 & 36.85 \\
E.T. C& 49.43 $\pm$ 1.07 &63.16  \\
\textbf{LensCraft (Ours)} & \textbf{1.66 $\pm$ 0.87} & \textbf{1.64} \\
\bottomrule
\end{tabular}
\end{table}

LensCraft achieves significantly lower inference time compared to both CCD and E.T., which enables real-time or near real-time camera trajectory generation. This efficiency stems from our lightweight transformer-based architecture and the use of a denoising masked autoencoder, which balances reconstruction robustness with computational frugality.

Despite operating under tighter resource constraints, LensCraft not only retains high output quality but also delivers substantial gains in responsiveness and scalability. These characteristics make it suitable for interactive and on-the-fly cinematography applications, where latency and resource usage are critical considerations.

\section{Qualitative Results}
\label{app: qualitative_result}
Given the inherent complexity of camera trajectories and the absence of a metric that can capture all aspects of cinematographic quality, complementing quantitative evaluations with qualitative assessments becomes essential. 
We conduct visual comparisons across three distinct scenarios to directly observe and evaluate how smooth and consistent the trajectories are, similar to the high level textual instructions we provided them. We present detailed analyses of these qualitative evaluations in the following subsections.

\subsection{Prompt only Results}
\begin{figure}
\centering
\includegraphics[width=1\linewidth]{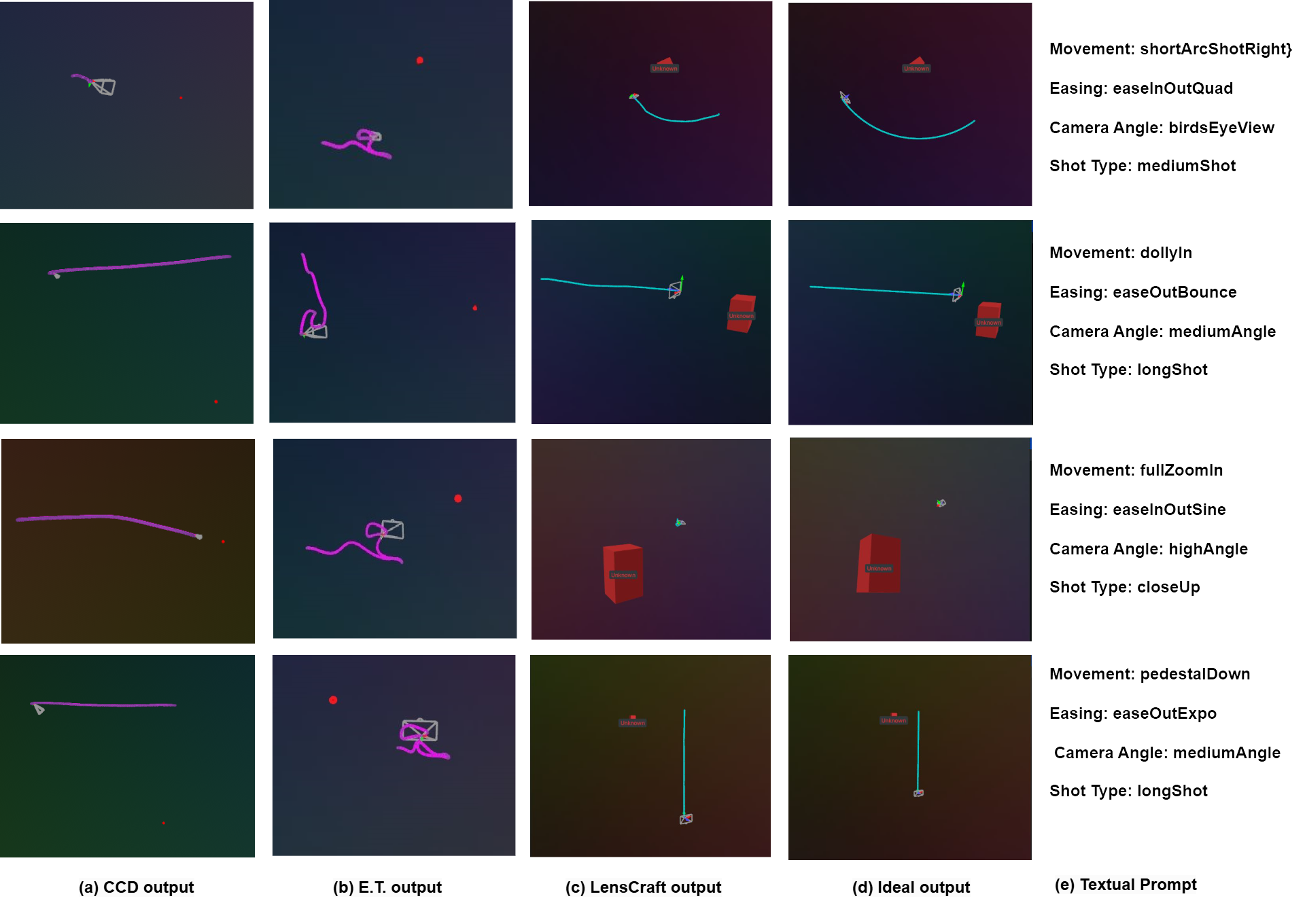}
\caption{Sample outputs comparing text-only inputs across models.}
\label{fig: result-prompt-only}
\end{figure}

In first scenario, we compare our model's outputs with SOTA models, E.T. \cite{courant2025exceptional} and CCD \cite{jiang2024cinematographic} in the text-to-camera trajectory generation task. This comparison specifically examines the model's ability to understand and execute professional cinematographic instructions expressed in natural language.
For a fair comparison, we evaluate all models using only textual inputs. Figure \ref{fig: result-prompt-only} presents sample outputs from this comparison.

As observed in Figure \ref{fig: result-prompt-only}, our model shows superior performance in interpreting user instructions, generating trajectories that more closely align with the ideal camera movements for given textual prompts. While all models utilize pretrained CLIP embeddings, our approach in splitting (with our high level representation) appears to better leverage CLIP's representational space for distinguishing and encoding camera movements. This results in trajectories that more accurately reflect the intended cinematographic instructions and maintain closer adherence to professional filming principles.

\subsection{Source trajectory Conditioning}
\begin{figure}
\centering
\includegraphics[width=1\linewidth]{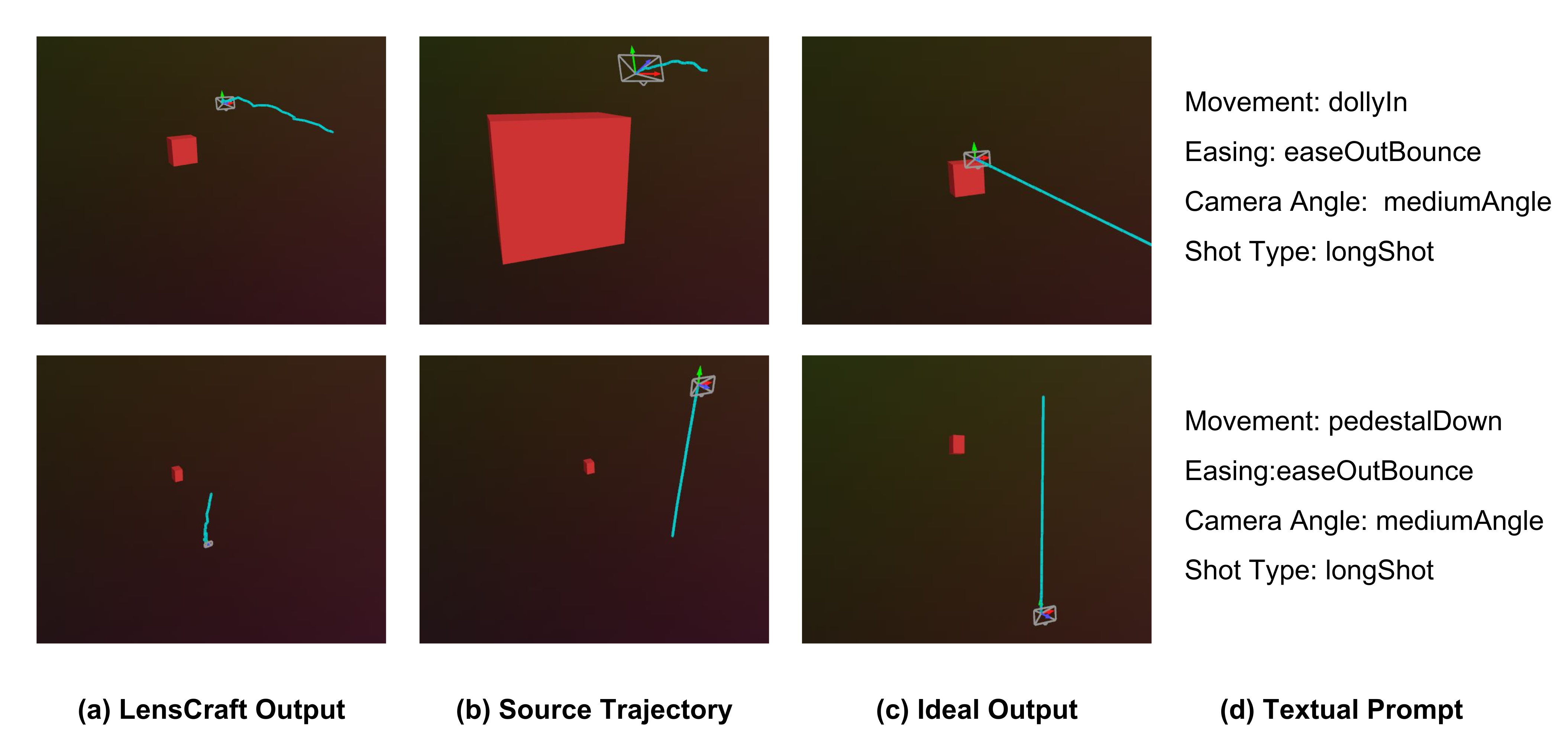}
\caption{LensCraft outputs in conditioning on reference trajectories and textual prompts.}
\label{fig: result-traj}
\end{figure}

In our second experiment, we evaluate the model's performance in conditioning on reference video camera trajectories. For this evaluation, we input the reference camera trajectory to the Encoder and provide textual instructions to the intermediate layer. Figure \ref{fig: result-traj} demonstrates the model's outputs under these conditions.

Figure \ref{fig: result-traj} demonstrates our model's robust capability to effectively integrate and reconcile features from reference trajectories with user-provided textual instructions. In the top two samples, where the input text and reference trajectory are from the same source, the model generates trajectories that closely match both inputs as expected. More interestingly, in the bottom two examples, where we intentionally input conflicting text prompts and reference trajectories, the model adopts a balanced approach, producing outputs that lean more towards satisfying the explicit textual instructions. This behavior aligns with our design objectives, as it suggests that when users explicitly specify certain attributes in their prompts that differ from the reference trajectory, the model interprets this as an intention to transfer only the non-conflicting characteristics from the reference while prioritizing the explicitly stated requirements.

\subsection{KeyFrame Conditioning}
\begin{figure}
\centering
\includegraphics[width=1\linewidth]{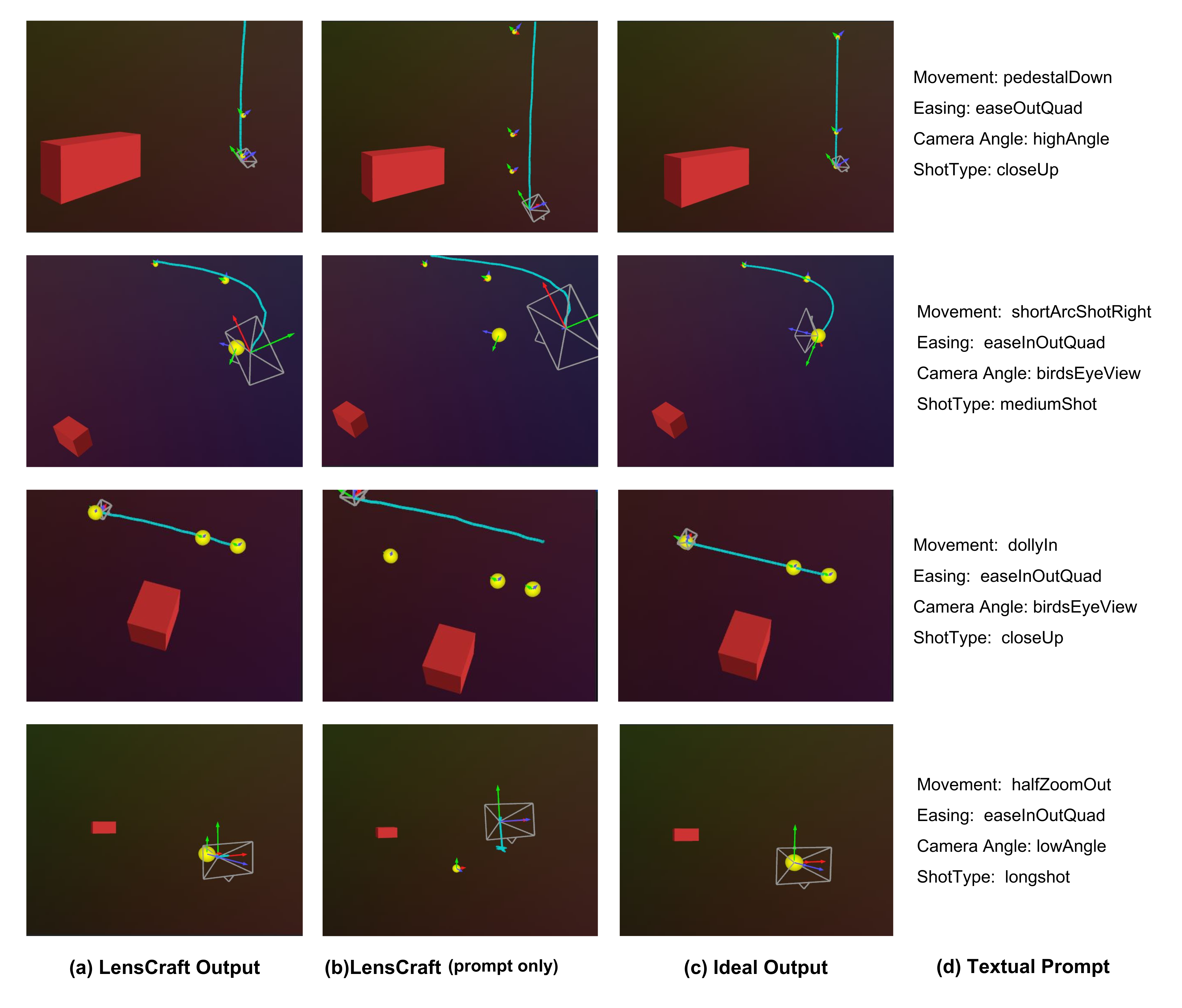}
\caption{LensCraft outputs with keyframe and textual prompt conditioning.}
\label{fig: result-keyframing}
\end{figure}

In our third experiment, we evaluate the model's performance in keyframe conditioning. We input user-specified keyframes to the Encoder as a masked reference trajectory. Figure \ref{fig: result-keyframing} presents the results of this evaluation.

The yellow points in the visualization represent the constrained keyframes, and as demonstrated in the fourth column, our model successfully navigates through these points whether they appear at the beginning, middle, or end of the sequence. The generated trajectories show strong similarity to the ideal trajectories, effectively interpolating between keyframes while maintaining natural camera motion. Based on these comprehensive evaluations across different conditioning scenarios, we could say our proposed solution demonstrates robust performance and offers a viable approach to the camera trajectory generation problem.

\section{Broader Impact}
\label{app: impacts}
The proposed LensCraft framework introduces a paradigm shift in automated cinematography by operationalizing high-level creative directives into semantically aligned, volumetrically aware camera trajectories. This contribution holds substantial implications for both research and practice in computer graphics, virtual production, and intelligent content generation.

By abstracting cinematographic reasoning into a multi-modal transformer architecture and leveraging a rigorously constructed simulation-driven dataset, LensCraft enhances accessibility to professional-grade camera planning tools. This advancement facilitates high-fidelity visual storytelling for creators with limited technical expertise or production resources, including independent filmmakers, digital educators, and interactive media designers.

Beyond content democratization, LensCraft's design fosters reproducibility and extensibility through open-source dissemination of its dataset, simulation environment, and model weights. Such transparency promotes cross-disciplinary adoption and downstream applications in domains like virtual reality, robotics perception, and synthetic data generation for computer vision tasks.

Overall, LensCraft is a step forward in making cinematic-quality tools accessible, supporting a more diverse and creative ecosystem for digital media production.


  

\end{document}